\documentclass[12pt]{article}
\usepackage{amsmath}
\usepackage{amssymb}
\usepackage{setspace}
\usepackage{geometry}
\usepackage[natbibapa]{apacite} 
\usepackage{amsthm}
\newtheorem{theorem}{Theorem}

\begin{document}

\title{Capacity Constraints in Principal-Agent Problems}
\author{Aubrey Clark\footnote{This is the second chapter from my PhD dissertation. I am grateful to my PhD advisors Eric Maskin, Oliver Hart, and Mihai Manea. For comments on various versions of the paper we thank Gabriel Carroll, Ben H{\'e}bert, Ryota Iijima, Divya Kirti, Jonathan Libgober, Xiaosheng Mu, Jeff Picel, Gleb Romanyuk, Jann Spiess, Eduard Talam{\`a}s, Yao Zeng, and audiences from the Theory and Contracts lunches at Harvard and MIT. Email: aubs.bc@gmail.com}}
\date{May, 2017}
\maketitle

\begin{abstract}
Adding a capacity constraint to a hidden-action principal-agent problem results in the same set of Pareto optimal contracts as the unconstrained problem where output is scaled down by a constant factor. This scaling factor is increasing in the agent's capacity to exert effort.
\end{abstract}

\section{Introduction}

A widespread economic situation is the relation between a principal and an agent (see \cite{Arrow1986}). In the hidden action version of this problem, an agent takes a costly action that affects the welfare of a principal. The relation between a physician and her patient, potential parties of a tort, investor and entrepreneur, and landlord and tenant are prominent examples. A feature of this relation is the set of actions available to the agent---the ways the agent can influence the distribution over outcomes. This set is determined by constraints on the environment (e.g. constraints on resources, technological know-how, etc). For example, a physician is constrained by the state of medical knowledge at the time she must make a diagnosis as well as by the amount of time and hospital resources she can devote to her patient. How should incentives change for the agent with only half the time to perform her task or after new technology expands the actions available to her?

This paper addresses an aspect of this question. The agent's action is modeled as a distribution over an unknown state whose realization determines outcomes; attached to each distribution is the cost the agent incurs if he wishes to implement it; and the agent’s choice is constrained to a feasible set of such distributions. We focus on modifying this feasible set via a capacity constraint capping the maximum cost of effort. Our main result applies when the principal and agent are risk-neutral. Here we obtain a relationship between optimal contracts with and without the capacity constraint.

To understand the result, imagine an optimal contract for which the agent’s capacity constraint binds---that is, if the cap on effort cost were slightly relaxed the agent would change action. What relation does this contract bear to the optimal contract obtained after doing away with the capacity constraint? What we show is that imposing the capacity constraint is equivalent to scaling down output by some factor $\alpha^*$ in $[0,1]$. That is, we find $\alpha^*$ such that given a mapping from output to optimal contracts for the problem without a capacity constraint, an optimal contract in the problem with a capacity constraint may be obtained by applying this mapping to a fraction $\alpha^*$ of output. We show that $\alpha^*$ is increasing in the agent’s capacity. Thus, expanding the set of feasible distributions by removing a capacity constraint increases the strength of incentives.

The proof goes as follows. Starting at 1, imagine scaling down output by a factor $\alpha$ until the capacity constraint stops binding for at least one optimal contract. Denote this value $\alpha^*$. For all $\alpha$ in $[\alpha^*, 1]$ we show that the set of optimal contracts is the same and for $\alpha^*$ there is an optimal contract that remains optimal after doing away with the capacity constraint.

Subject to optimal contracts existing, we do not impose a particular structure on the sets of feasible distributions and feasible contracts; the agent’s cost function need only be continuous. This accommodates a wide range of restrictions on allowable contracts (e.g., liability limits, monotonicity, inability to discriminate between collections of states, etc.) and a wide range of restrictions on how the agent can influence the distribution over states.

We may apply our result to the literature on capital structure (e.g., \cite{ModiglianiMiller1958}; see \cite{Myers2001} for a review article). A sector of that literature seeks to derive capital structure from agency between investors and those running firms (e.g., \cite{JensenMeckling1976}). A common result is that optimality requires the firm to raise capital exclusively through the sale of debts (see \cite{Innes1990}). Puzzling, then, that firms sell equity, too. An application of our result resolves this puzzle: the optimal contract under a capacity constraint is debt and a fraction $1 - \alpha^*$ of equity, the remaining equity kept inside the firm to motivate those running it.

\section{Framework}
Output depends on the realization of one among a finite set of states $\Omega$ as described by an \textit{output function} $y : \Omega \to \mathbb{R}$. An agent has the ability to choose the distribution $p \in \Delta(\Omega)$. A principal hires the agent to choose this distribution but cannot verify the choice. The state is verifiable and the agent is given incentives to choose one distribution or another with a \textit{contract}
\[
b : \Omega \to \mathbb{R}.
\]

A \textit{profile} $(b, p)$ consists of a contract $b$ and distribution $p$. The agent’s preferences over profiles $(b, p)$ are represented by the expected value of contract payments less a distribution cost:
\[
E_p[b] - c(p) = \sum_\omega p(\omega) b(\omega) - c(p).
\]

The function $c : \Delta(\Omega) \to \mathbb{R}$ is the agent’s \textit{cost function}. The principal’s preferences over profiles $(b, p)$ are represented by the expected value of output less contract payments:
\[
E_p[y - b] = \sum_\omega p(\omega)(y(\omega) - b(\omega)) .
\]

Contracts are restricted to belong to a set $\mathcal{B}$ of \textit{feasible contracts}. Distributions are restricted to belong to a set $\mathcal{D}$ of \textit{feasible distributions}.

The agent faces a \textit{capacity constraint}, which means that included in the restrictions that define the set of feasible distributions $\mathcal{D}$ is the condition that the cost of each is no more than the agent’s capacity $k \in \mathbb{R}$:
\[
c(p) \leq k.
\]

A profile $(b, p)$ is \textit{feasible} if the contract $b$ is feasible, the distribution $p$ is feasible, and this profile is the agent’s most preferred profile among those consisting of contract $b$ and any feasible distribution. Profile $(b, p)$ may be \textit{improved upon} if there exists a feasible profile that makes either the agent or principal better off without making the other worse off. \textit{Pareto optimal profiles} are those that are feasible and cannot be improved upon.

\section{Examples}

Many principal-agent setups fit this framework.

\subsection{The textbook principal-agent problem}

In the textbook model, the set of feasible distributions $\mathcal{D}$ is parameterized by effort $e \in \mathbb{R}$, and the distribution over output corresponding to effort $e$ is denoted
\[
p(\cdot|e).
\]
The agent’s cost function maps effort $e$ to its cost $c(p(\cdot|e))$. The set of feasible contracts $\mathcal{B}$ is often defined by liability limits.

\subsection{Information acquisition}

One may be more explicit about how the agent influences the distribution over states. In problems of information acquisition, a state $\omega$ is taken to be a decision $d$ chosen by the agent and a state of nature $\theta$ that the agent cannot influence but can acquire information about. The agent faces a statistical decision problem. He acquires information about the state of nature $\theta$ with an \textit{experiment} $(X, \{q(\cdot|\theta)\}_{\theta \in \Theta})$, that consists of a set of signals $X$ and a collection of probability distributions $\{q(\cdot|\theta)\}_{\theta \in \Theta} \subseteq \Delta(X)$ over signals, and he chooses a decision rule $f : X \to \Delta(D)$, mapping signals to randomizations over decisions. Given a prior $\pi$ on the states of nature, an experiment and decision rule induce a distribution $p(\cdot|\pi, X, \{q(\cdot|\theta)\}_{\theta \in \Theta}, f) \in \Delta(\Omega)
$ over states.

\subsection{Multiple periods}

The distribution over states may arise from the agent’s behavior in multiple periods. One could take the states $\Omega$ as the set of terminal histories $(\theta_1, \theta_2, \ldots, \theta_T)$ where each $\theta_t, t = 1, 2, \ldots, T$, is information revealed to the agent at time $t$. For each history $(\theta_1, \ldots, \theta_{t-1})$, the agent chooses a distribution $q(\cdot|\theta_1, \ldots, \theta_{t-1})$ over $\theta_t$. This induces a distribution $p(\cdot|q) \in \Delta(\Omega)$.


\section{Result}

Consider a perturbation of our model, parametrized by $\alpha \in [0, 1]$, in which output $y$ in the principal’s objective function is replaced by $\alpha y$ (the feasible contracts, distributions, and profiles stay the same). Denote the set of Pareto optimal profiles to the perturbed problem by $\mathcal{P}(\alpha)$. Further denote by $\mathcal{P}(\alpha, r)$ the subset of $\mathcal{P}(\alpha)$ in which the agent receives the lowest level of utility among the elements of $\mathcal{P}(\alpha)$ that is at least equal to $r$.

\begin{theorem}\label{thm:main} Let $(b, p) \in \mathcal{P}(1, r)$ and define $\alpha^* \in [0, 1]$ as
\[
\alpha^* = \sup \left\{ \alpha \in [0, 1] : c(p') < k \text{ for all } (b', p') \in \mathcal{P}(\alpha, E_p[b] - c(p)) \right\}.
\]
Then, for each $\alpha \in [\alpha^*, 1]$,
\[
\mathcal{P}(\alpha, E_p[b] - c(p)) \supseteq \mathcal{P}(1, r),
\]
and, conversely, if $(b_\alpha, p_\alpha) \in \mathcal{P}(\alpha, E_p[b] - c(p))$ is such that $c(p_\alpha) = k$, then
\[
(b_\alpha, p_\alpha) \in \mathcal{P}(1, r).
\]

If the agent’s cost function is continuous, then there is a Pareto optimal profile in $\mathcal{P}(\alpha^*, E_p[b] - c(p))$ that solves the problem obtained by removing the capacity constraint from the perturbed problem for $\alpha^*$.
\end{theorem}

The point is that all solutions $\mathcal{P}(1, r)$ to the unperturbed problem (a problem in which the capacity constraint might bind) may be obtained as solutions $\mathcal{P}(\alpha^*, E_p[b] - c(p))$ to the perturbed problem for $\alpha^*$ (and for at least one solution of this problem the capacity constraint does not bind).


\section{Proof}

Let $(b, p) \in \mathcal{P}(1, r)$, define $\alpha^*$ as in the statement of the theorem, and let $\alpha \in [\alpha^*, 1]$. 

The proof has two steps. For step 1 we will show that if $(b_\alpha, p_\alpha) \in \mathcal{P}(\alpha, E_p[b] - c(p))$, then the following inequalities hold:

\begin{equation}
E_p[y] - E_{p_\alpha}[y] \geq E_p[b] - E_{p_\alpha}[b] \geq \alpha(E_p[y] - E_{p_\alpha}[y]) \geq 0, \label{eq:2.1}
\end{equation}
\begin{equation}
c(p) - c(p_\alpha) \geq E_p[b] - E_{p_\alpha}[b]. \label{eq:2.2}
\end{equation}

For step 2, we will show these inequalies imply the conclusion of the theorem: 

\[
\mathcal{P}(\alpha, E_p[b] - c(p)) \supseteq \mathcal{P}(1, r),
\]
and, conversely, if $(b_\alpha, p_\alpha) \in \mathcal{P}(\alpha, E_p[b] - c(p))$ is such that $c(p_\alpha) = k$, then
\[
(b_\alpha, p_\alpha) \in \mathcal{P}(1, r).
\]

\paragraph{Step 1.}

To obtain the inequalities in \eqref{eq:2.1}, note $E_p[y - b] \geq E_p[y - b_\alpha]$ and $E_{p_\alpha}[y - b_\alpha] \geq E_p[y - b]$. The first and second inequalities of \eqref{eq:2.1} are rearrangements of these; and, from this, the last emerges because $\alpha \in [0, 1]$. Inequality \eqref{eq:2.2} follows from rearrangement of the participation constraint $E_{p_\alpha}[b] - c(p_\alpha) \geq E_p[b] - c(p)$.

\paragraph{Step 2.} By the definition of $\alpha^*$ we  have $c(p_\alpha) = k$. The inequalities in \eqref{eq:2.2} then give $c(p) = k$ and $E_p[b] = E_{p_\alpha}[b]$. The inequalities in \eqref{eq:2.1} then give $E_p[y] = E_{p_\alpha}[y]$. Therefore $(b, p) \in \mathcal{P}(\alpha, E_p[b] - c(p))$ (i.e., $\mathcal{P}(\alpha, E_p[b] - c(p)) \supseteq \mathcal{P}(1, r)$) and $(b_\alpha, p_\alpha) \in \mathcal{P}(1, r)$.

Finally, we show there is a Pareto optimal profile in $\mathcal{P}(\alpha^*, E_p[b] - c(p))$ that solves the problem obtained by removing the capacity constraint from the perturbed problem for $\alpha^*$. Let $\alpha_1, \alpha_2, \ldots$ be a sequence in $[0, \alpha^*)$ converging to $\alpha^*$ and let $(b_1, p_1), (b_2, p_2), \ldots$ be a sequence of profiles with $(b_i, p_i) \in \mathcal{P}(\alpha_i, E_p[b] - c(p))$, $i = 1, 2, \ldots$. Since the agent’s cost function is continuous, the set of feasible profiles yielding the agent at least utility $E_p[b] - c(p)$ is compact. Therefore, there is a subsequence $(b_{n_1}, p_{n_1}), (b_{n_2}, p_{n_2}), \ldots$ converging to some $(b', p')$. Continuity of the principal’s objective function implies $(b', p') \in \mathcal{P}(\alpha^*, E_p[b] - c(p))$. Continuity of the agent’s cost function together with the fact that $c(p_i) < k$ (because $\alpha_i < \alpha^*$ for $i = 1, 2, \ldots$) implies $(b', p')$ solves the problem obtained by removing the capacity constraint from the perturbed problem for $\alpha^*$. \hfill $\Box$

\section{Application to a Theory of Capital Structure}

Suppose that in the absence of a capacity constraint all Pareto optimal contracts correspond to debt contracts
\[
y(\omega) - b(\omega) = \min\{y(\omega), F\},
\]
for some $F \geq 0$, so that $b(\omega) = \max\{0, y(\omega) - F\}$.

By Theorem \ref{thm:main}, if the agent’s cost function is continuous, then provided $\mathcal{P}(1, r)$ is nonempty, at least one of its elements $(b, p)$ has the form $b(\omega) = \max\{0, \alpha^* y(\omega) - F\} = \alpha^* \max\{0, y(\omega) - F / \alpha^*\}$.

The corresponding payoff for the principal is a debt of $F / \alpha^*$ and a fraction $1 - \alpha^*$ of the remaining equity:
\[
y(\omega) - b(\omega) = \min\{y(\omega), F / \alpha^*\} + (1 - \alpha^*) \max\{0, y(\omega) - F / \alpha^*\}.
\]

By its definition, $\alpha^*$ increases with agent capacity. Thus, as agent capacity increases, more debt and less equity is issued.

Note that our result does not assume a particular form for the feasible contracts. For example, we could apply our result in a setting where contracts are restricted to be monotonically increasing in output with a slope less than one. Our result also does not assume the form of the feasible set of distributions, so we could also apply it if the set of distributions were restricted to satisfying a monotone likelihood ratio property. \cite{Innes1990} uses both assumptions in deriving debt contracts as optimal, and thus we can apply our result to his model.

\subsection{Live-or-Die Contracts}

Dropping contract monotonicity, \cite{Innes1990} finds that the resulting optimal contract corresponds to a live-or-die security:
\[
y(\omega) - b(\omega) = 
\begin{cases} 
y(\omega) & \text{if } y(\omega) < l, \\
0 & \text{if } y(\omega) \geq l.
\end{cases}
\]
Then
\[
b(\omega) = 
\begin{cases} 
0 & \text{if } y(\omega) < l, \\
y(\omega) & \text{if } y(\omega) \geq l.
\end{cases}
\]

The argument applied to the debt and equity case applies here too. In the capacity-constrained problem, if output is $\alpha^*$ of itself, then (i) there is an optimal contract such that the capacity constraint does not bind and (ii) this contract is optimal for the original problem (the problem with unaltered output). Therefore the agent’s contract will have the form shown here but with output replaced by a fraction $\alpha^*$ of itself. The resulting contract form for the principal is
\[
y(\omega) - b(\omega) = 
\begin{cases} 
y(\omega) & \text{if } y(\omega) < l, \\
(1 - \alpha^*) y(\omega) & \text{if } y(\omega) \geq l.
\end{cases}
\]

Under a capacity constraint, the principal retains the fraction $1 - \alpha^*$ of output in the “die” region of the live-or-die contract since he only needs to exhaust the agent’s capacity, and this can be done with the fraction $\alpha^*$ of output.

\section{Discussion}

The capital structure irrelevance theorems of \cite{ModiglianiMiller1958} state that the market value of a firm is invariant to capital structure because buyers of securities may borrow or lend, making it as if they bought from a firm with any other capital structure.

Many theories now exist to explain why some capital structures are preferred to others (see \cite{Myers2001} for a review). These theories relax the assumptions of the Modigliani-Miller theorems: the absence of borrowing constraints and taxes, that only debt and equity may be issued, or that the cash flows of the firm are independent from its capital structure.

My results relate to the agency theory of capital structure in which capital structure determines the incentives of the owners of a firm and therefore its cash flows. This point of view was first taken in \cite{JensenMeckling1976}, who considered a firm’s incentives for issuing debt versus equity. Debt is good because it leaves intact entrepreneur incentives whenever firm profits exceed the face value of debt. Equity is bad because it scales down the entrepreneur’s incentives: an entrepreneur that sells a 5 percent equity stake will only exert effort to the point where the marginal cost of additional effort is equal to the marginal benefit of an additional 95 cents, not the full 100 cents that their additional effort generates. After observing this, Jensen and Meckling ask why firms issue equity at all?

\begin{quote}
An ingenious entrepreneur eager to expand, has open to him the opportunity to design a whole hierarchy of fixed claims on assets and earnings, with premiums paid for different levels of risk. Why don’t we observe large corporations individually owned with a tiny fraction of the capital supplied by the entrepreneur in return for 100 percent of the equity and the rest simply borrowed?
\end{quote}

Their answer was that too much debt encourages the entrepreneur to take greater risks in order to obtain the high payoffs where their incentives lie (the alternative being the minimum payoff implied by limited liability). Thus, equity is issued because at some point the harm caused from the risk-taking induced by additional debt outweighs the benefits of retaining entrepreneurial effort incentives.

My results suggest an alternative reason why firms issue equity: why leave the entrepreneur with 100 percent equity when a lesser fraction will do? That is, if the entrepreneur is capacity constrained, then selling a certain amount of equity does not change their incentives and so is the best way to initially raise capital. At some point, the capacity constraint stops binding. If at this point additional capital is necessary, then it becomes optimal to issue debt.

\section{Conclusion}

We have shown that in solving for optimal incentives, imposing a capacity constraint on the agent is equivalent to scaling down output by a factor $\alpha^* \in [0, 1]$ and that this factor is increasing in the agent’s capacity. When applied to models that yield debt as the optimal contract, a capacity constraint implies that the optimal contract becomes a debt and a fraction $1 - \alpha^*$ of equity. This contributes to the literature that seeks to explain capital structure from a principal-agent relationship between investors and those running the firm.

It would be interesting to find an analogue of our results for when the agent is risk-averse or when the feasible set of actions for the agent changes in a way other than through a capacity constraint.
\newpage
\appendix

\section{Appendix}

\subsection{Risk-Aversion}

We have been unable to find an analogue of our result for when the agent is risk-averse. If the principal is risk-neutral and the agent is risk-averse, the problem in our proof is that \eqref{eq:2.2} now involves a Bernoulli utility function $u$, and therefore we cannot conclude (i) $E_p[u(b)] - E_{p_\alpha}[u(b_\alpha)] = 0$ \text{ from \eqref{eq:2.1}}, nor (ii) that (i) implies: $E_p[u(b)] - E_{p_\alpha}[u(b_\alpha)] = 0$.

The problem may be further seen by considering first-order conditions for the simple capacity-constrained problem:
\[
\max_{b, p} E_p[y - b]
\]
\text{ such that } 

\[c(p) \leq k,\]
\begin{equation}
E_p[u(b)] - c(p) \geq E_{p'}[u(b)] - c(p') \text{ for all } p' \text{ such that } c(p') \leq k, \label{A.1}
\end{equation}
and
\[
E_p[u(b)] - c(p) \geq 0.
\]

Here the set of feasible contracts is unrestricted and the set of feasible distributions are those satisfying the capacity constraint $c(p) \leq k$. The agent’s program may be written as:
\[
\max_p E_p[u(b)] - c(p)
\]
such that $p(\omega) \geq 0$ for all $\omega \in \Omega$, $\sum_{\omega \in \Omega} p(\omega) = 1$, and $c(p) \leq k$. 

If the agent’s cost function is convex and the optimal distribution always assigns positive probability to each state, then a necessary and sufficient condition for $p$ to be an optimal contract is:
\begin{equation}
u(b(\omega)) - \frac{\partial c(p)}{\partial p(\omega)} = \rho + \mu \frac{\partial c(p)}{\partial p(\omega)}, \label{A.2}
\end{equation}
where $\rho$ and $\mu$ are Lagrange multipliers on the constraints $\sum_{\omega \in \Omega} p(\omega) = 1$ and $c(p) \leq k$. 

If the agent’s cost function is strictly convex, then one may show that the program obtained after replacing the incentive constraint \eqref{A.1} by the first-order condition \eqref{A.2} and making $\mu$ a choice variable satisfies a qualification constraint, which implies that any solution (in which $p$ assigns positive probability to each state) satisfies first-order condition

\[
y(\omega) - b(\omega) = \tau + \delta \frac{\partial c(p)}{\partial p(\omega)} - (\mu + 1) \sum_{\omega' \in \Omega} \phi[\omega'] \frac{\partial^2 c(p)}{\partial p(\omega)\partial p(\omega')} + \zeta \left( u(b(\omega)) - \frac{\partial c(p)}{\partial p(\omega)} \right),
\]
\[
-p(\omega) = \phi[\omega] u'(b(\omega)) + \zeta p(\omega) u'(b(\omega)),
\]
\[
0 = - \sum_{\omega \in \Omega} \phi[\omega] \frac{\partial c(p)}{\partial p(\omega)},
\]
where $\tau$ and $\delta$ are Lagrange multipliers on the adding-up and capacity constraints $\sum_{\omega \in \Omega} p(\omega) = 1$ and $c(p) \leq k$, $\phi[\omega]$ is a Lagrange multiplier on the agent’s first-order condition \eqref{A.2}, and $\zeta$ is a Lagrange multiplier on the participation constraint $E_p[b] - c(p) \geq 0$. 

Rearranging the first conditions and using the agent’s first-order condition yields

\[
b(\omega) + \frac{\delta + \mu}{1 + \mu} u(b(\omega)) = y(\omega) - A - B \sum_{\omega' \in \Omega} p(\omega) \left( \frac{1}{u'(b(\omega))} + \zeta \right) \frac{\partial^2 c(p)}{\partial p(\omega) \partial p(\omega')},
\]
for constants $A$ and $B$. If the agent is risk-neutral, then this may be written as:
\[
b(\omega) = \frac{1 + \mu}{1 + \mu + \mu \delta} y(\omega) - A - B \sum_{\omega' \in \Omega} p(\omega) \left( \frac{1}{u'(b(\omega))} + \zeta \right) \frac{\partial^2 c(p)}{\partial p(\omega) \partial p(\omega')},
\]
which suggests our result. Such a representation does not seem to be possible when the agent is risk-averse.
    
\subsection{Discounting}

Discounting does not pose a problem when contract payoffs arise at a single date. In this case, we can define a new cost function for the agent equal to his old cost function divided by his discount factor. However, when payoffs arrive at different dates our proof fails for similar reasons as when the agent is averse to risk. 

Suppose that payments can be made at two dates: date 0 and date 1. Suppose that the principal and agent discount payments with the principal’s discount factor denoted by $\delta_P \in [0,1]$ and the agent’s discount factor denoted by $\delta_A \in [0,1]$. Output and contracts are now functions of dates as well as states: $y, b : \{0,1\} \times \Omega \to \mathbb{R}$. 

The principal’s utility from profile $(b, p)$ is now:
\[
E_p(\omega) \left[ \sum_{t \in \{0,1\}} \delta_P^t (y(t, \omega) - b(t, \omega)) \right],
\]
and the agent’s utility is:
\[
E_p(\omega) \left[ \sum_{t \in \{0,1\}} \delta_A^t b(t, \omega) \right] - c(p).
\]

Then, following the proof of the case without discounting, we obtain the inequalities:
\[
E_p(\omega) \left[ \sum_{t \in \{0,1\}} \delta_P^t y(t, \omega) \right] - E_{p_\alpha}(\omega) \left[ \sum_{t \in \{0,1\}} \delta_P^t y(t, \omega) \right] 
\]
\[
\geq E_p(\omega) \left[ \sum_{t \in \{0,1\}} \delta_A^t b(t, \omega) \right] - E_{p_\alpha}(\omega) \left[ \sum_{t \in \{0,1\}} \delta_A^t b(t, \omega) \right]
\]
\[
\geq \alpha \left( E_p(\omega) \left[ \sum_{t \in \{0,1\}} \delta_P^t y(t, \omega) \right] - E_{p_\alpha}(\omega) \left[ \sum_{t \in \{0,1\}} \delta_P^t y(t, \omega) \right] \right) \geq 0,
\]
\[
c(p) - c(p_\alpha) \geq E_p(\omega) \left[ \sum_{t \in \{0,1\}} \delta_A^t b(t, \omega) \right] - E_{p_\alpha}(\omega) \left[ \sum_{t \in \{0,1\}} \delta_A^t b_\alpha(t, \omega) \right]. 
\]

As under risk-aversion, we cannot conclude that the value of discounted contract payments for the agent is the same under both contracts, nor that this implies their discounted value is equal for the principal.


\newpage
\begin{thebibliography}{}

\bibitem[Arrow(1986)]{Arrow1986}
Arrow, K. J. (1986). Agency and the market. \textit{Handbook of Mathematical Economics}, 3:1183--1195.

\bibitem[Clark(2017)]{Clark2017}
Clark, A. (2017). \textit{Principal-Agent Problems and Experimentation}. Ph.D. thesis, Harvard University.

\bibitem[Innes(1990)]{Innes1990}
Innes, R. D. (1990). Limited liability and incentive contracting with ex-ante action choices. \textit{Journal of Economic Theory}, 52:45--67.

\bibitem[Jensen and Meckling(1976)]{JensenMeckling1976}
Jensen, M. C., \& Meckling, W. H. (1976). Theory of the firm: Managerial behavior, agency costs and ownership structure. \textit{Journal of Financial Economics}, 3(4):305--360.

\bibitem[Modigliani and Miller(1958)]{ModiglianiMiller1958}
Modigliani, F., \& Miller, M. H. (1958). The cost of capital, corporation finance and the theory of investment. \textit{The American Economic Review}, 48:261--297.

\bibitem[Myers(2001)]{Myers2001}
Myers, S. C. (2001). Capital structure. \textit{The Journal of Economic Perspectives}, 15(2):81--102.

\end{thebibliography}
\end{document}